\documentclass[long, final]{jobim}

\usepackage[utf8x]{inputenc} 

\usepackage{tikz}
\usetikzlibrary{arrows}
\usetikzlibrary{patterns}
\usetikzlibrary{decorations.pathreplacing}
\usepackage[normalem]{ulem}

\usepackage{multirow}
\usepackage{filecontents}

\title{LRez: C++ API and toolkit for analyzing and managing Linked-Reads data}

\author{Pierre \textsc{Morisse}\inst{1} \and Claire \textsc{Lemaitre}\inst{1} \and Fabrice \textsc{Legeai}\inst{1},\inst{2}}

\institute{
 Univ Rennes, Inria, CNRS, IRISA, 35000, Rennes, France 
 \and
 INRAE, UMR 1349 INRAE/Agrocampus Ouest/Université Rennes 1, Institut de Génétique, Environnement et Protection des Plantes (IGEPP), F-35653 Le Rheu, France
}

\corresponding{pierre.morisse@inria.fr}

\abstract{%
	Linked-Reads technologies, such as 10x Genomics, combine both the high-quality and low cost of short-reads sequencing and a long-range information, through the use of barcodes able to tag reads which originate from a common long DNA fragment. This technology has been employed in a broad range of applications including assembly or phasing of genomes, and structural variant calling. However, to date, no tool or API dedicated to the manipulation of Linked-Reads data exist. We introduce LRez, a C++ API and toolkit which allows easy management of Linked-Reads data. LRez includes various functionalities, for computing number of common barcodes between genomic regions, extracting barcodes from BAM files, as well as indexing and querying both BAM and FASTQ files to quickly fetch reads or alignments sharing one or multiple barcodes. LRez can thus be used in a broad range of applications requiring barcode processing, in order to improve their performances. LRez is implemented in C++, supported on Linux platforms, and available under AGPL-3.0 License at \href{https://github.com/morispi/LRez}{https://github.com/morispi/LRez}.}

\begin{document}

\selectlanguage{english}

   \maketitle

\section{Introduction}

Linked-Reads technologies, popularized by 10x Genomics \cite{Zheng2016}, partition and tag high-molecular-weight DNA molecules with a barcode using a microfluidic device prior to classical short-read sequencing. This way, Linked-Reads manage to combine the high-quality of the short reads and a long-range information which can be inferred by determining the origin of the short-reads fragments with the help of the barcodes. Although 10x Genomics discontinued the sales of their product lines, other comparable technologies such as TELL-Seq \cite{Chen2020} and Haplotagging \cite{Meier2020} recently emerged. Moreover, large volumes of data were already produced and still need to be analyzed. \\ 
\indent To benefit from the Linked-Reads data, most methods first map all the reads against a reference genome, and then rely on the analysis of the barcode contents of genomic regions. However, despite the fact various tools libraries such as SAMtools \cite{Wysoker2009} and BamTools \cite{Barnett2011} tools are available for processing BAM files, to the best of our knowledge, no toolkit currently exists to manage Linked-Reads barcodes. LRez aims to address this issue, by providing a complete and easy to use API and suite of tools designed to serve that purpose.

\section{Features and methods}

\subsection{LRez toolkit}

The LRez toolkit provide end-users with a suite of utilities for manipulating Linked-Reads barcodes, both in FASTQ and BAM files. Table \ref{tab:toolkitCommands} provides a summary of currently available features of LRez, and some of these features are further described below.

\subsubsection{Index BAM \\ \\} 

\hspace*{.5em} The \texttt{index bam} command can be used to build two types of indexes. In each case, the index is represented as map, where the key is the barcode, recorded using 2 bits per nucleotide to mitigate memory consumption, and where the value is a list of elements depending on the index type. The first type of index records the offsets of the barcodes in the BAM file, allowing to retrieve all alignments involving reads with a given barcode. The second type of index records the genomic positions of the alignments of reads in which the barcode appears, in format chromosome:position. It can then be queried, for instance, to quickly retrieve the chromosomes on which a given barcoded molecule appears. During construction, BAM files are processed using a slightly altered version of BamTools, allowing to retrieve the offset of any given alignment in the BAM file. \\
\indent Moreover, the construction of both types of indexes supports additional parameters allowing to only consider barcodes from primary alignments and / or having a mapping quality higher than a given threshold. These options allow to restrict the barcodes of interest to the most valuable ones for further analysis, and thus reduce memory footprint and querying time.

\subsubsection{Index Fastq \\ \\}

\hspace*{.5em} Similarly to the \texttt{index bam} subcommand, LRez supports the indexing and querying of native Chromium 10x FASTQ files, or of any other barcoded FASTQ files, given the barcodes are reported in the read headers with the BX:Z tag. This barcode extraction can be performed using external tools such as Long Ranger basic. In the exact same way as for BAM files, \texttt{index fastq} stores all the offset positions of the reads in a optimized format, allowing to rapidly extract reads sharing one or multiple barcodes.


\subsubsection{Compare \\ \\}

\hspace*{.5em}  The \texttt{compare} command computes the number of common barcodes either between pairs of regions or between contig ends of the reference genome. The first feature takes as input a BAM file and a file containing the regions of interest (in format chromosome:startPosition-endPosition), and computes common barcodes counts between all possible pairs of regions. So far, it does not make use of the index and performs greedy extraction / comparison of the barcode contents of each region. It was implemented as so to avoid requiring the user to build the index, which can be time consuming, if only a few comparisons are required. The second feature takes as input the a BAM file, a contig of interest, as well as the length of the extremities to consider (1,000 by default). This time, the index is used, since extracting the barcodes from all contig ends would be extremely time consuming in the case of fragmented assemblies. The barcodes from the extremities of the contig of interest are extracted, and the index is queried in order to retrieve other contig ends in which these barcodes appear, and compute the numbers of common barcodes on-the-fly.

\begin{table}
\begin{center}
	\begin{tabular}{lcl}
		\hline
		\textbf{Command} & & \textbf{Description} \\
		\hline
		\multirow{2}{*}{compare} & & Compute the number of common barcodes between pairs \\
		& & of regions or between pairs of contig ends \\
		extract & &  Extract the barcodes from a given region of a BAM file \\
		\multirow{2}{*}{index bam} & & Index the BAM offsets or genomic positions of the \\
		& & barcodes contained in a BAM file \\
		\multirow{2}{*}{index fastq} & & Index by barcode the offsets of the sequences \\
		& & contained in a FASTQ file \\
		\multirow{2}{*}{query bam} & & Query the index to retrieve alignments in a BAM file \\
		& & given a barcode or list of barcodes \\
		\multirow{2}{*}{query fastq} & & Query the index to retrieve sequences in a FASTQ file \\
		& & given a barcode or list of barcodes \\
		\hline \\
	\end{tabular}
	\caption{LRez command-line toolkit features.}
	\label{tab:toolkitCommands}
\end{center}
\end{table} 

\subsubsection{Results \\ \\}

\hspace*{.5em} We tested the \texttt{index bam} subcommand on a 61 GB BAM file generated by the Genome In A Bottle project from \emph{H. sapiens} NA24385 reads. Index construction took an hour, and resulted in an 9 GB index. We then used the \texttt{query bam} subcommand to query the index with 100,000 random barcodes. Querying time per barcode mainly varied according to the number of alignments the barcode was involved in, and the number of jumps through the BAM file required. It thus ranged from 2 ms up to 5 seconds, for barcodes involved in as much as 10,000 alignments. \\
\indent We also tested the \texttt{compare} subcommand to compute the number of common barcodes between contig ends of the 1,000 largest contigs of a highly fragmented \emph{H. numata} assembly \cite{Jay}. After building the index, processing time per contig reached an average of 2 seconds. With 8 threads, the comparison between all contigs only required 4 minutes.

\subsection{LRez API}

The LRez API provides C++ programmers with tools allowing to efficiently analyze and manage Linked-Reads data. It is compiled as a shared .so library, helping its integration to external projects. Moreover, all functionalities are implemented in a thread-safe fashion. Available modules include \texttt{IndexManagementBam} and \texttt{IndexManagementFastq}, to build indexes from BAM and FASTQ files, \texttt{AlignmentsRetrieval}, and \texttt{ReadsRetrieval}, to query the indexes and retrieve alignments or reads tagged with a given barcode, and finally \texttt{BarcodesExtraction} and \texttt{BarcodesComparison}, to extract barcodes from given regions of a BAM file, and compute the number of common barcodes between pairs of regions or between contig ends. This latter feature is particularly interesting, in the sense that it has the potential to be directly applicable to scaffolding or gap-filling projects. \\

\indent The following example illustrates how to build the index from a BAM file, and how to use this index to compute the number of common barcodes between the ends of a given contig, and the ends of all other contigs:

\begin{verbatim}
	bool onlyIndexPrimary = false;
	unsigned minQuality = 0;
	BarcodesOffsetsIndex index;
	index = indexBarcodesOffsetsFromBam(bamFile,
	        onlyIndexPrimary, minQuality);
	unsigned extSize = 1000;
	compareContig(bamFile, index, "chr12", extSize);
\end{verbatim}

\section{Conclusion}


LRez provides both a toolkit and a easy to use C++ API, allowing to deal with Linked-Reads data and offering various functionalities. We thus believe it might help both end-users and programmers alike, and be easily integrated to external projects, including phasing, structural variant discovery or genome scaffolding. Currently, it is already used in the structural variant calling tool LEVIATHAN (\href{https://github.com/morispi/LEVIATHAN}{https://github.com/morispi/LEVIATHAN}), where its indexing and querying features are used to efficiently compute the numbers of common barcodes between all possible pairs of regions of the reference genome. Furthermore, FASTQ indexing and querying as well as contig comparison functionalities are used by the scaffolding / gap-filling tool MTG-Link (\href{https://github.com/anne-gcd/MTG-Link}{https://github.com/anne-gcd/MTG-Link}), to mitigate resource consumption. Since new bioinformatics tool are still needed to deal with worthwhile Linked-Reads from the most recent technologies (\cite{Chen2020}, \cite{Meier2020}), we believe LRez has the potential to be used in a wide variety of other applications.

\section{Acknowledgements}

This project has received funding from the French ANR ANR-18-CE02-0019 Supergene grant.  

\bibliography{bibliography.bib} 
\end{document}